\documentclass[prd,showpacs,preprintnumbers,amsmath,amssymb]{revtex4}

\usepackage{graphicx}
\usepackage{dcolumn}
\usepackage{bm}

\usepackage{color}

\begin{document}

\title{Schwinger Effect from Near-extremal Black Holes in (A)dS Space}

\author{Chiang-Mei Chen} \email{cmchen@phy.ncu.edu.tw}
\affiliation{Department of Physics, National Central University, Chungli 32001, Taiwan}
\affiliation{Center for High Energy and High Field Physics (CHiP), National Central University, Chungli 32001, Taiwan}

\author{Sang Pyo Kim}\email{sangkim@kunsan.ac.kr}
\affiliation{Department of Physics, Kunsan National University, Kunsan 54150, Korea}
\affiliation{Institute of Theoretical Physics, Chinese Academy of Sciences, Beijing 100190, China}

\date{\today}

\begin{abstract}
We study the Schwinger effect in near-extremal Reissner-Nordstr\"{o}m (RN) black holes with electric and/or magnetic charges in the (Anti-) de Sitter (AdS) space. The formula for the Schwinger effect takes a universal form for near-extremal black holes with the near-horizon geometry of ${\rm AdS}_2 \times S^2$ and with the proper radii for the ${\rm AdS}_2$ space and the two-sphere $S^2$, regardless of the asymptotically flat or (A)dS space. The asymptotic AdS boundary enhances and the dS boundary suppresses the Schwinger effect and the small radius of the AdS (dS) space reinforces the enhancement and suppression.
\end{abstract}


\maketitle

\section{Introduction}

Quantum fluctuations spontaneously create particle pairs from the vacuum unless there exists an external mechanism preventing annihilation of pairs. The event horizon of a black hole casually separates pairs into the interior and the exterior and the black hole emits all species of particles~\cite{Hawking:1974sw}. A sufficiently strong electric field separates charged pairs and accelerates them in opposite directions, the so-called Schwinger pair production~\cite{Sauter:1932ab, Schwinger:1951nm}. The Hawking radiation and Schwinger effect are consequence of the nonperturbative effects of quantum field theory.

The Hawking radiation and Schwinger effect both act as the emission channels of charged pairs from charged black holes. Hawking radiation obeying the Bose-Einstein or Fermi-Dirac distribution with the Hawking temperature emits charges with a chemical potential of Coulomb potential on the horizon. Though the Hawking radiation from (near-) extremal black holes is exponentially suppressed due to vanishing Hawking temperature, the Schwinger mechanism triggers the emission of charges. The Schwinger effect has been extensively studied for (near-) extremal Reissner-Nordstr\"{o}m (RN) black holes~\cite{Chen:2012zn, Chen:2014yfa}, Kerr-Newman (KN) black holes~\cite{Chen:2016caa}, dyonic KN black holes~\cite{Chen:2017mnm} (for review and references, see Ref.~\cite{Kim:2019joy}). The Schwinger fermion production has also been studied for charged black holes in the (A)dS space~\cite{Belgiorno:2007va, Belgiorno:2008mx, Belgiorno:2009pq}.

Primordial black holes have been studied from cosmology and astrophysics~\cite{Hawking:1971ei, Carr:1974nx, Carr:1975qj}. It was argued that dynonic extremal black holes are stable against Hawking radiation~\cite{Gibbons:1976sm}. Black holes of planck scale are stable quantum mechanically and can be a candidate for dark matter~\cite{Aharonov:1987tp}. The Schwinger effect from (near-) extremal RN and dyonic RN black holes in (Anti-) de Sitter space is interesting theoretically and cosmologically since the asymptotic (A)dS boundary affects the near-horizon geometry of (near-) extremal black holes~\cite{Montero:2019ekk}. In particular, the expanding early universe described by a dS space drastically changes the property and evolution of charged black holes.

We study the effect of the asymptotic (A)dS boundary on Schwinger pair production in (near-) extremal RN black holes. The extremal RN-(A)dS black holes are obtained by degenerating the event horizon and the causal horizon while keeping the cosmological horizon separately from the black hole horizon. The near-horizon region of (near-) extremal RN-(A)dS black holes has the geometry of ${\rm AdS}_2 \times S^2$, whose symmetry allows explicit solutions of a charged field in terms of confluent hypergeometric or hypergeometric functions. The asymptotic (A)dS boundary makes the effective radii of the ${\rm AdS}_2$ and the two-sphere for a charged field from that of (near-) extremal RN black holes in the asymptotically flat space~\cite{Chen:2012zn, Chen:2014yfa}. The formula for the mean number for pair production has a universal structure in terms of the effective temperature for charge acceleration and Hawking temperature with modifications due to the effect of the asymptotic (A)dS boundary. We then investigate the pair production of dyons with electric and magnetic charges from (near-) extremal dyonic RN-(A)dS black holes. The formula for the mean number of dyon pairs exhibits the universal structure.

The organization of this paper is as follows. In Sec.~\ref{rn-ads}, we study the Schwinger effect in (near-) extremal RN-(A)dS black holes. The near-horizon geometry is still ${\rm AdS}_2 \times S^2$, whose effective radii for the ${\rm AdS}_2$ space and $S^2$ are modified by the asymptotic radius of the (A)dS space. The Schwinger effect has a universal structure in terms of the effective temperature for charges and the Hawking temperature, and is suppressed by the dS boundary but enhanced by the AdS boundary. In Sec.~\ref{dyon-rn-ads}, the Schwinger effect is studied in (near-) extremal dyonic black holes in the (A)dS space. The universal structure for the mean number of dyon pairs shows the Schwinger effect in ${\rm AdS}_2$ and quantum electrodynamics (QED) effect of the electromagnetic field in a Rindler space for the surface gravity of the event horizon. The effective temperature is determined by the Unruh temperature for the acceleration of dyons in the electromagnetic field. The Schwinger effect is suppressed in the dS space but enhanced in the AdS space. In Sec.~\ref{con}, we discuss cosmological and astrophysical implications of the effect of the asymptotic (A)dS boundary on the Schwinger effect.

\section{(Near-) Extremal RN-(A)dS Black Holes} \label{rn-ads}

The action for RN-(A)dS black holes is given by
\begin{eqnarray}
S = \int d^4x \sqrt{-g} \left( R \pm \frac6{L^2} - F_{\mu \nu} F^{\mu \nu} \right).
\end{eqnarray}
where a cosmological constant $\Lambda = \mp 3/L^2$ (the upper/lower sign) here and hereafter corresponds to the AdS/dS space, respectively.
The metric for an RN-(A)dS black holes is
\begin{eqnarray}
ds^2 &=& - f(r) dt^2 + \frac{dr^2}{f(r)} + r^2 d\Omega_2^2, \qquad f(r) = 1 - \frac{2M}{r} + \frac{Q^2}{r^2} \pm \frac{r^2}{L^2},
\nonumber\\
A &=& \frac{Q}{r} dt, \label{rn-ads}
\end{eqnarray}
where $M$ and $Q$ are the mass and charge of the black hole and $L$ is the (A)dS radius. There are two positive roots of $f(r)$, in general, associated to the causal horizon $r_-$ and event horizon $r_+$, and for the dS case a third positive root corresponds to the cosmological horizon $r_{\rm C}$~\cite{Romans:1991nq}. Out of three horizons of RN-(A)ds black holes~(\ref{rn-ads}), the causal horizon $r_-$ and the event horizon $r_+$ are made degenerate, $M = M_0, r_+ = r_- = r_0$, to yield an extremal RN-(A)dS black holes, where
\begin{eqnarray} \label{M0}
M_0 = \frac{L (3 \pm \delta) \sqrt{\delta}}{3 \sqrt6}, \qquad r_0 = \frac{L \sqrt{\delta}}{\sqrt6}, \qquad \delta = \pm \left( \sqrt{1 \pm 12 Q^2/L^2} - 1 \right).
\end{eqnarray}
The cosmological horizon remains outside of the event horizon ($r_{\rm C} > r_0$). By elongating the radial coordinate $r = r_0 + \epsilon \rho$ and reducing the time coordinate, $t = \tau/\epsilon$, one has the near-horizon geometry of extremal RN-(A)dS
\begin{eqnarray}
ds^2 = - \frac{\rho^2}{R_\mathrm{AdS}^2} d\tau^2 + \frac{ R_\mathrm{AdS}^2}{\rho^2} d\rho^2 + R_\mathrm{S}^2 d\Omega_2^2, \qquad A = - \frac{Q}{R_\mathrm{S}^2} \rho \, d\tau,
\end{eqnarray}
possessing a structure of AdS$_2 \times S^2$ with radii
\begin{eqnarray} \label{RAdS}
R_\mathrm{AdS}^2 = \frac{L^2 \delta}{6 (1 \pm \delta)}, \qquad R_\mathrm{S}^2 = r_0^2 = \frac{L^2 \delta}{6}.
\end{eqnarray}

From Eq.~(\ref{M0}) we find another expression
\begin{eqnarray}\label{RS-M0-AdS}
R_\mathrm{S} = \frac{L}{\sqrt{6}} \Biggl[ \Bigl( \sqrt{1 + \frac{27}{2} \frac{M_0^2}{L^2}} + \sqrt{\frac{27}{2}} \frac{M_0}{L} \Bigr)^{1/3}
- \Bigl( \sqrt{1 + \frac{27}{2} \frac{M_0^2}{L^2}} + \sqrt{\frac{27}{2}} \frac{M_0}{L} \Bigr)^{-1/3} \Biggr],
\end{eqnarray}
for the AdS space
\begin{eqnarray}\label{RS-M0-dS}
R_\mathrm{S} &=& -i \frac{L}{\sqrt{6}} \Biggl[ \Bigl( \sqrt{1 - \frac{27}{2} \frac{M_0^2}{L^2}} + i \sqrt{\frac{27}{2}} \frac{M_0}{L} \Bigr)^{1/3}
- \Bigl( \sqrt{1 - \frac{27}{2} \frac{M_0^2}{L^2}} + i \sqrt{\frac{27}{2}} \frac{M_0}{L}\Bigr)^{-1/3} \Biggr]
\nonumber\\
&=& \frac{2 L}{\sqrt{6}} \sin\frac{\vartheta}3, \qquad \vartheta = \sin^{-1} \sqrt{\frac{27}{2}} \frac{M_0}{L}, \qquad 0 \le \vartheta \le \pi/2,
\end{eqnarray}
for the dS space. The black hole radius takes the large $L$ limit
\begin{eqnarray}
R_\mathrm{S} = M_0 \Bigl[ 1 \mp 2 \Bigl( \frac{M_0}{L} \Bigr)^2 + 12 \Bigl( \frac{M_0}{L} \Bigr)^4 \mp 96 \Bigl( \frac{M_0}{L} \Bigr)^6  + {\cal O} \Bigl( \frac{M_0}{L} \Bigr)^8 \Bigr].
\end{eqnarray}
The radius of $\mathrm{AdS}_2$  and the charge are
\begin{eqnarray}
R_\mathrm{AdS}^2 = \frac{R_\mathrm{S}^2}{1 \pm 6 \bigl( \frac{R_\mathrm{S}}{L} \bigr)^2}, \qquad Q = R_\mathrm{S} \sqrt{1 \pm 3 \Bigl( \frac{R_\mathrm{S}}{L} \Bigr)^2}.
\end{eqnarray}

We consider a charged scalar governed by the Klein-Gordon equation,
\begin{eqnarray}
(\nabla_\mu - i q A_\mu) (\nabla^\mu - i q A^\mu) \Phi - m^2 \Phi = 0,
\end{eqnarray}
whose solution
\begin{eqnarray}
\Phi(\tau, \rho, \theta, \varphi) = \mathrm{e}^{-i \omega \tau + i n \varphi} R(\rho) S(\theta)
\end{eqnarray}
is separated into the angular and radial equations as
\begin{eqnarray}
\frac1{\sin\theta} \partial_\theta (\sin\theta \partial_\theta S) - \left( \frac{n^2}{\sin^2\theta} - l (l + 1) \right) S &=& 0,
\\
\partial_\rho (\rho^2 \partial_\rho R) + \left[ \frac{R_\mathrm{AdS}^4}{R_\mathrm{S}^4} \frac{(q Q \rho - \omega R_\mathrm{S}^2)^2}{\rho^2} - R_\mathrm{AdS}^2 m^2 - \frac{R_\mathrm{AdS}^2}{R_\mathrm{S}^2} l (l + 1) \right] R &=& 0.
\end{eqnarray}
By introducing an effective mass and $R_\mathrm{AdS}^2$-rescaled electric force
\begin{eqnarray}
\bar m^2 = m^2 + \frac{l (l + 1)}{R_\mathrm{S}^2} + \frac{1/4}{R_\mathrm{AdS}^2}, \qquad \kappa = R_\mathrm{AdS}^2 \frac{q Q}{R_\mathrm{S}^2},
\end{eqnarray}
the violation of the Breitenlohler-Freedman (BF) bound in AdS$_2$ space leading to the Schwinger pair production takes the form~~\cite{Pioline:2005pf, Kim:2008xv,Cai:2014qba}
\begin{eqnarray} \label{BFbound}
\mu^2 = \kappa^2 - R_\mathrm{AdS}^2 \bar m^2 \geq 0.
\end{eqnarray}
The BF bound ($\mu^2 < 0$) of AdS$_2$ space guarantees the stability against the pair production. The solution to the radial equation is in terms of the Whittaker function~\cite{Chen:2012zn}
\begin{eqnarray}
R = c_1 M_{i \kappa, i \mu}(z) + c_2 M_{i \kappa, -i \mu}(z), \qquad z = i \frac{2 \omega R_\mathrm{AdS}^2}{\rho}.
\end{eqnarray}

\begin{figure}[h]
\includegraphics[width=0.45\linewidth,height=0.35\textwidth]{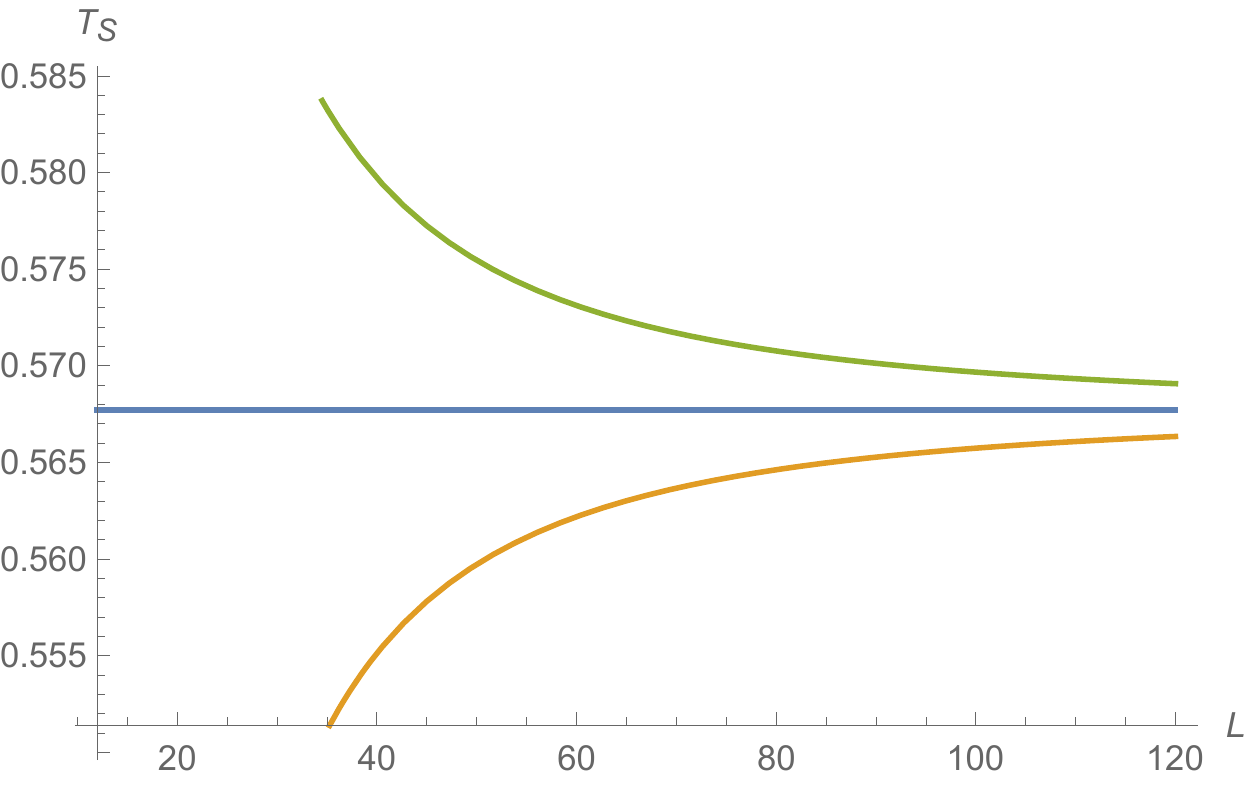}
\hfill
\includegraphics[width=0.45\linewidth,height=0.35\textwidth]{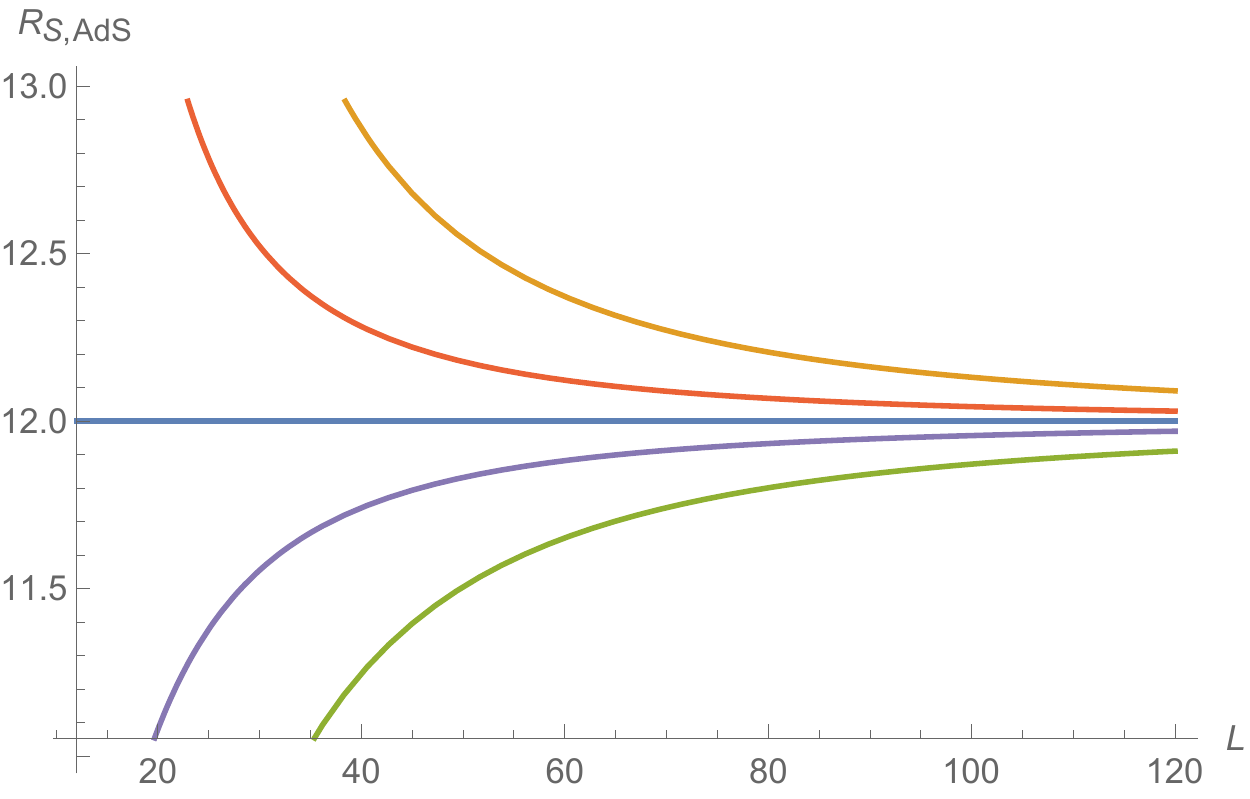}
\caption{\label{T-S-R-S-C}
[Left panel] The effective temperature $T_\mathrm{S}$ against $L$ with $Q = \sqrt{12}, q = \pi, m = 1, l = 0$ fixed. The green (upper) and yellow (lower) curves are the effective temperature for extremal RN black holes in the AdS and dS spaces, respectively, while the blue horizontal line is that in an asymptotically flat space. [Right panel] The AdS$_2$ radius $R_\mathrm{AdS}$ and the two-sphere radius $R_\mathrm{S}$ against $L$ with $Q = \sqrt{12}, q = \pi, m = 1, l = 0$ fixed. The horizontal line denotes both $R_\mathrm{AdS}$ and $R_\mathrm{S}$ in the asymptotically flat space. The yellow (uppermost) curve denotes $R_\mathrm{AdS}$ and the orange curve above the horizontal line denotes $R_\mathrm{S}$ in the dS space while the violet curve below the horizontal line denotes $R_\mathrm{S}$ and the green curve (lowermost) denotes $R_\mathrm{AdS}$ in the AdS space.}
\end{figure}

According to our earlier result in Ref.~\cite{Chen:2012zn}, the mean number of produced pairs is
\begin{eqnarray}
\mathcal{N} = \frac{\sinh(2 \pi \mu)}{\cosh(\pi \kappa + \pi \mu)} \exp(\pi \mu - \pi \kappa) = \frac{\mathrm{e}^{-2 \pi (\kappa - \mu)} - \mathrm{e}^{-2 \pi (\kappa + \mu)}}{1 + \mathrm{e}^{-2 \pi (\kappa + \mu)}}. \label{ex-sch pair}
\end{eqnarray}
The mean number~(\ref{ex-sch pair}) of charges from the extremal black hole has a thermal interpretation
\begin{eqnarray}
\mathcal{N} = \frac{\mathrm{e}^{- \bar m/T_\mathrm{S}} - \mathrm{e}^{- \bar m/\bar T_\mathrm{S}}}{1 + \mathrm{e}^{- \bar m/\bar T_\mathrm{S}}},
\end{eqnarray}
in terms of the effective temperature
\begin{eqnarray}
T_\mathrm{S} &=& \frac{\bar m}{2 \pi (\kappa - \mu)} = T_\mathrm{U} + \sqrt{T_\mathrm{ U}^2 - \frac1{4 \pi^2 R_\mathrm{AdS}^2}}, \qquad T_\mathrm{ U} = \frac{\kappa /R_\mathrm{AdS}^2}{2 \pi \bar m},
\nonumber\\
\bar T_\mathrm{ S} &=& \frac{\bar m}{2 \pi (\kappa + \mu)} = T_\mathrm{U} - \sqrt{T_\mathrm{U}^2 - \frac1{4 \pi^2 R_\mathrm{AdS}^2}}. \label{eff tem}
\end{eqnarray}
Note that the effective temperature~(\ref{eff tem}) is the same as that of charges in a uniform electric field in AdS$_2$ space, in which $T_\mathrm{U}$ is the Unruh temperature for the accelerated charge by the electric field and the second term in the square root is related to the BF bound~\cite{Pioline:2005pf, Kim:2008xv}, which corresponds to the Gibbons-Hawking temperature square in dS$_2$ space~\cite{Cai:2014qba}. The factor of two, i.e. $T_\mathrm{S} = 2 T_\mathrm{U}$, in the Minkowski spacetime limit ($R_\mathrm{AdS} = \infty$) is an ultra-relativistic feature of the Schwinger effect. In general, $T_\mathrm{S}$ for the AdS space is larger than, and $T_\mathrm{S}$ for the dS space is the smaller than that for the asymptotically flat space, as shown in Fig.~\ref{T-S-R-S-C}. The $T_\mathrm{S}$ for the (A)dS space approaches that for the asymptotically flat space. The reason for this is that the AdS$_2$ radius $R_\mathrm{AdS}$ and the two-sphere radius $R_\mathrm{S}$ for the dS space are larger than those for the AdS space, see Fig.~\ref{T-S-R-S-C}, and thereby the electric field on the horizon is weaker, the Unruh temperature lower for the dS space than those for the AdS space. The asymptotic (A)dS space drastically changes the effective temperature: the higher temperature for the AdS space enhances the Schwinger effect while the lower temperature for the dS space suppresses the Schwinger effect. Remarkably the dS space increases the AdS$_2$ radius $R_\mathrm{AdS}$ for BF bound, which implies larger extremal RN black holes stable against the Schwinger effect than the asymptotically flat space and opens possibility for large extremal primordial black holes.

\begin{figure}
\includegraphics[width=0.45\linewidth,height=0.35\textwidth]{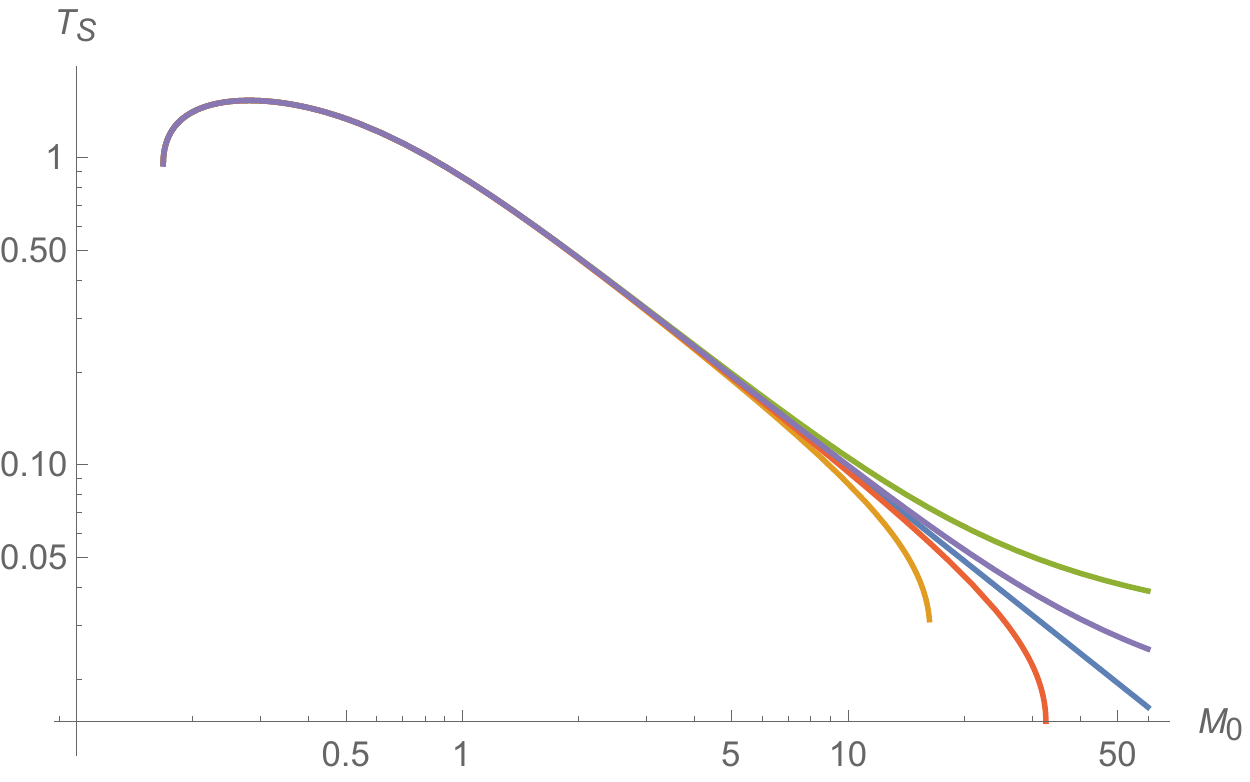}
\hfill
\includegraphics[width=0.45\linewidth,height=0.35\textwidth]{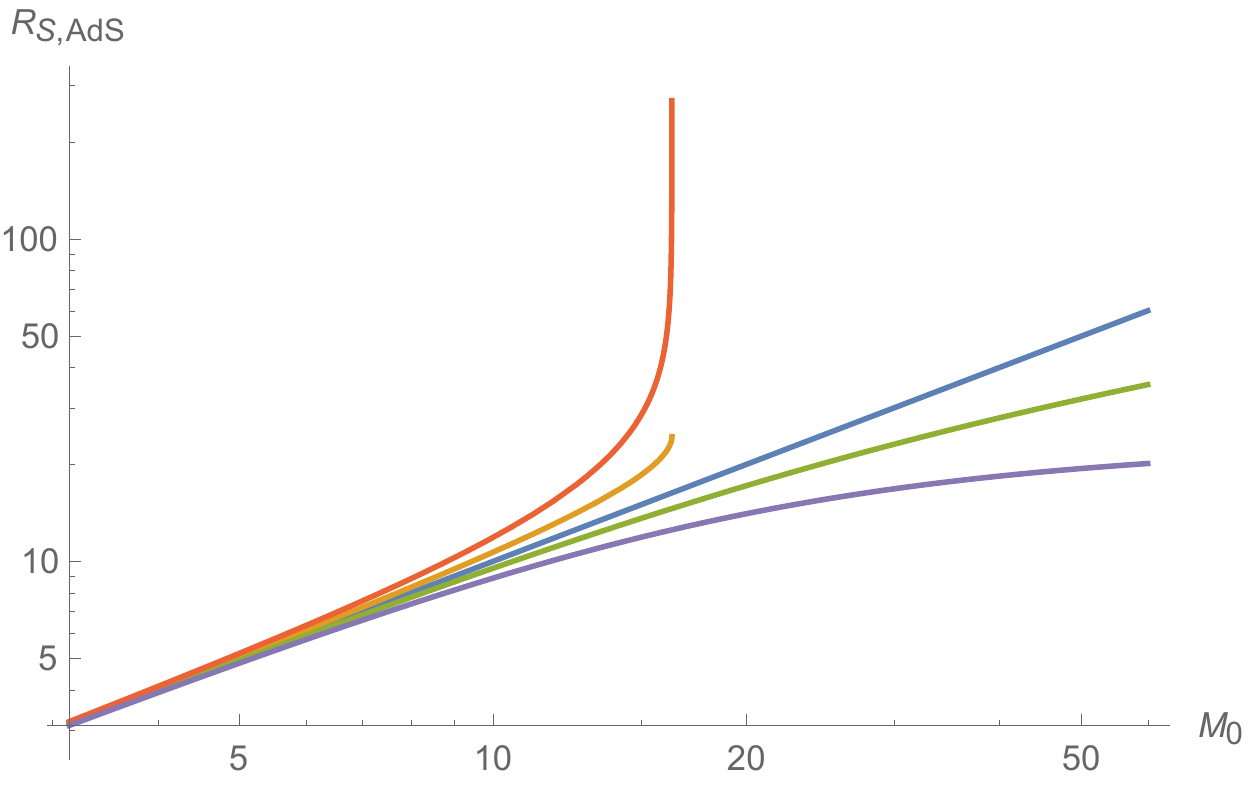}
\caption{\label{T-U-R-S-C2}
[Left panel] The effective temperature $T_\mathrm{S}$ against $M_0$ with $q = \pi, m = 1, l = 0$ fixed. The blue (middle) curve is the effective temperature in the asymptotically flat space $(L = \infty)$. The green (upper most) and purple (above the blue) curves are $T_\mathrm{S}$ with  $L = 60$ and $L = 120$ respectively in the AdS space while the red (below the blue) and yellow (bottom most) curves are $T_\mathrm{S}$ with $L = 120$ and $L = 60$ respectively in the dS space. [Right panel] The $\mathrm{AdS}_2$ radius $R_\mathrm{AdS}$ and the two-sphere radius $R_\mathrm{S}$ against $M_0$ with $L = 60, q = \pi, m = 1, l = 0$ fixed. The blue curve is $R_\mathrm{AdS} = R_\mathrm{S}$ in the asymptotically flat space. The red (upper most) curve is $R_\mathrm{AdS}$ and the yellow (above the blue) curve is $R_\mathrm{S}$ in the dS space while the green (below the blue) curve is $R_\mathrm{S}$ and the purple (bottom most) curve is $R_\mathrm{AdS}$ in the AdS space.}
\end{figure}

The dependence of $T_\mathrm{S}$ and radii $R_\mathrm{S}, R_\mathrm{AdS}$ with respect to $M_0$ is given in~Fig~\ref{T-U-R-S-C2}. The effect of cosmological constant becomes more significant, enhanced in AdS and suppressed in dS, when $M_0$ is larger. The small $M_0$ threshold in $T_\mathrm{S}$ reflects the BF bound in~(\ref{BFbound}). The Schwinger effect is actually in accord with the unstable modes of AdS$_2$ in near horizon geometry. For the asymptotically dS spacetimes, there is an upper bound of $M_0$ such that the Schwinger mechanism terminated. This corresponds to the limit when the black hole horizon $R_\mathrm{S}$ approaches the cosmological horizon, namely $\delta = 1$ or $\vartheta = \pi/2$ in~(\ref{RS-M0-dS}), and then the $R_\mathrm{AdS}$ diverges.

\medskip
The near-horizon geometry of near-extremal RN black holes is obtained by scaling coordinates $r_{\pm} = r_0 \pm \epsilon B$ and parameterizing the mass
\begin{eqnarray} \label{nearM}
M = M_0 + (\epsilon B)^2 \frac{R_{\rm S}}{2 R_{\rm AdS}^2}.
\end{eqnarray}
The metric and gauge field for the near-extremal black holes are
\begin{eqnarray}
ds^2 = - \frac{\rho^2 - B^2}{R_\mathrm{AdS}^2} d\tau^2 + \frac{ R_\mathrm{AdS}^2}{\rho^2 - B^2} d\rho^2 + R_\mathrm{S}^2 d\Omega_2^2,
\qquad A = - \frac{Q}{R_\mathrm{S}^2} \rho \, d\tau. \label{near RN-ads}
\end{eqnarray}
The Hawking temperature and the chemical potential on the horizon in the metric~(\ref{near RN-ads}) are then given by
\begin{eqnarray}
T_{\rm H} = \frac{B}{2 \pi R_{\rm AdS}^2}, \qquad \Phi_{\rm H} = - \frac{QB}{R_{\rm S}^2},
\end{eqnarray}
while the Hawking temperature in the original metric is $\epsilon T_{\rm H}$ and suppressed by small $\epsilon$ due to the rescaling of time. The radial equation of charged scalar field is
\begin{eqnarray}
\partial_\rho \bigl[ (\rho^2 - B^2) \partial_\rho R \bigr] + \left[ \frac{R_\mathrm{AdS}^4}{R_\mathrm{S}^4} \frac{(q Q \rho - \omega R_\mathrm{S}^2)^2}{\rho^2 - B^2} - R_\mathrm{AdS}^2 m^2 - \frac{R_\mathrm{AdS}^2}{R_\mathrm{S}^2} l (l + 1) \right] R &=& 0.
\end{eqnarray}
The solutions to the radial equation are found in terms of the hypergeometric functions, which lead to the mean number for pair production
\begin{eqnarray}
\mathcal{N} = \frac{\sinh(2 \pi \mu) \sinh(\pi \tilde{\kappa} - \pi \kappa)}{\cosh(\pi \kappa + \pi \mu) \cosh(\pi \tilde{\kappa} - \pi \mu)} = \frac{\mathrm{e}^{-2 \pi (\kappa - \mu)} - \mathrm{e}^{-2 \pi (\kappa + \mu)}}{1 + \mathrm{e}^{-2 \pi (\kappa + \mu)}} \times  \frac{1 - \mathrm{e}^{-2 \pi (\tilde{\kappa} - \kappa)}}{1 + \mathrm{e}^{-2 \pi (\tilde{\kappa} - \mu)}}, \label{near-sch pair}
\end{eqnarray}
where
\begin{eqnarray}
\kappa = \frac{R_{\rm AdS}^2}{R_{\rm S}^2} q Q = - \frac{q \Phi_{\rm H}}{2 \pi T_{\rm H}}, \qquad \tilde{\kappa} = \frac{\omega R_{\rm AdS}^2}{B} = \frac{\omega}{ 2 \pi T_{\rm H}}.
\end{eqnarray}
The mean number has the following thermal interpretation~\cite{Kim:2015kna}
\begin{eqnarray}
\mathcal{N} = \underbrace{\Biggl( \frac{\mathrm{e}^{- \bar m/T_\mathrm{S}} - \mathrm{e}^{- \bar m/\bar T_\mathrm{S}}}{1 + \mathrm{e}^{- \bar m/\bar T_\mathrm{S}}} \Biggr)}_\textrm{Schwinger effect in AdS$_2$} \times \Biggl\{ \mathrm{e}^{\bar m/T_\mathrm{S}} \underbrace{\Biggl( \mathrm{e}^{- \bar m/T_\mathrm{S}} \frac{1 - \mathrm{e}^{-(\omega + q \Phi_\mathrm{H})/T_\mathrm{H}}}{1+  \mathrm{e}^{- \bar m/T_\mathrm{S}} \mathrm{e}^{-(\omega + q \Phi_\mathrm{H})/T_\mathrm{H}}} \Biggr)}_\textrm{Schwinger effect in Rindler$_2$} \Biggr\}. \label{rn-ther-int}
\end{eqnarray}
The first parenthesis is the Schwinger effect for an extremal black hole. The second curly bracket is expected since the Hawking temperature for a near-extremal black hole does not completely vanish, though small, and the Hawking radiation and the Schwinger effect is intertwined. The surface gravity of the event horizon determines the acceleration of the two-dimensional Rindler space and the effective temperature is the QED effect in the electric field of charged black hole.

\section{(Near-) Extremal Dyonic RN-(A)dS Black Holes} \label{dyon-rn-ads}

The second model is the dyonic RN-(A)dS black holes with the electric and magnetic charge $Q$ and $P$, which have the metric and the potentials for $Q$ and $P$
\begin{eqnarray}
ds^2 &=& - f(r) dt^2 + \frac{dr^2}{f(r)} + r^2 d\Omega_2^2, \qquad f(r) = 1 - \frac{2M}{r} + \frac{Q^2 + P^2}{r^2} \pm \frac{r^2}{L^2},
\nonumber\\
A &=& \frac{Q}{r} dt + P (\cos\theta \mp 1) d\varphi, \qquad \bar A = \frac{P}{r} dt - Q (\cos\theta \mp 1) d\varphi,
\end{eqnarray}
where the upper(lower) sign in $f(r)$ is for the AdS (dS) space. The magnetic monopole induces a string-like singularity causing diﬀerent choices of the gauge potential: the upper sign is regular in $0 \le \theta < \pi/2$, and the lower sign in $\pi/2 < \theta \le \pi$. The field strengths of $A$ and $\bar A$ are Hodge dual to each other. The extremal condition, $M_0$, and the radius of degenerated horizon, $r_0$, are given in~(\ref{M0}), with modification of
\begin{eqnarray}
\delta = \pm \bigl( \sqrt{1 \pm 12 (Q^2 + P^2)/L^2} -1 \bigr). \label{dyon-del}
\end{eqnarray}
Note that parameters for extremal dyonic RN black holes are obtained by replacing $Q^2+ P^2$ for $Q^2$ for extremal RN black holes. By stretching the radial coordinate $r = r_0 + \epsilon \rho$ but squeezing the time coordinate $t = \tau/\epsilon$ and taking near extremal condition~(\ref{nearM}), we obtain the near-horizon geometry of near extremal dyonic RN-(A)dS
\begin{eqnarray}
ds^2 &=& - \frac{\rho^2 - B^2}{R_\mathrm{AdS}^2} d\tau^2 + \frac{R_\mathrm{AdS}^2}{\rho^2 - B^2} d\rho^2 + R_\mathrm{S}^2 d\Omega_2^2,
\nonumber\\
A &=& - \frac{Q}{R_\mathrm{S}^2} \rho \, d\tau + P (\cos\theta \mp 1) d\varphi, \qquad \bar A = - \frac{P}{R_\mathrm{S}^2} \rho \, d\tau - Q (\cos\theta \mp 1) d\varphi,
\end{eqnarray}
where the radii $R_\mathrm{AdS}$ and $R_\mathrm{S}$ are given in Eq.~(\ref{RAdS}) with $\delta$ in Eq.~(\ref{dyon-del}). The electric and magnetic charges $q$ and $p$ of emitted particles are coupled to the potentials $A^\mu$ and $\bar A^\mu$ as
\begin{eqnarray}
(\nabla_\mu - i q A_\mu - i p \bar A_\mu) (\nabla^\mu - i q A^\mu - i p \bar A^\mu) \Phi - m^2 \Phi = 0,
\end{eqnarray}
whose wave function takes the form
\begin{eqnarray}
\Phi(\tau, \rho, \theta, \varphi) = \mathrm{e}^{-i \omega \tau + i [n \mp (q P - p Q)] \varphi} R(\rho) S(\theta),
\end{eqnarray}
The solution separates into the angular and the radial parts
\begin{eqnarray}
\frac1{\sin\theta} \partial_\theta (\sin\theta \partial_\theta S) - \left( \frac{[n - (q P - p Q) \cos\theta ]^2}{\sin^2\theta} - \lambda \right) S &=& 0,
\\
\partial_\rho \left[ (\rho^2 - B^2) \partial_\rho R \right] + \left[ \frac{R_\mathrm{AdS}^4}{R_\mathrm{S}^4} \frac{[ (q Q + p P) \rho - \omega R_\mathrm{S}^2]^2}{\rho^2 - B^2} - R_\mathrm{AdS}^2 m^2 - \frac{R_\mathrm{AdS}^2}{R_\mathrm{S}^2} \lambda \right] R &=& 0.
\end{eqnarray}
The result is analog with the electric black hole with the modifications
\begin{eqnarray} \label{dkappa}
\kappa = \frac{R_\mathrm{AdS}^2}{R_\mathrm{S}^2} (q Q + p P) = - \frac{q \Phi_\mathrm{H} + p \bar \Phi_\mathrm{H}}{2 \pi T_\mathrm{H}}, \qquad \bar \Phi_\mathrm{H} = - \frac{P B}{R_\mathrm{S}^2}.
\end{eqnarray}

The mean number generalizes the formula with zero angular momentum in Ref.~\cite{Chen:2017mnm} to the (A)dS boundary
\begin{eqnarray}
\mathcal{N} = \underbrace{\Biggl( \frac{\mathrm{e}^{- \bar m/T_\mathrm{S}} - \mathrm{e}^{- \bar m/\bar T_\mathrm{S}}}{1 + \mathrm{e}^{- \bar m/\bar T_\mathrm{S}}} \Biggr)}_\textrm{Schwinger effect in AdS$_2$} \times \Biggl\{ \mathrm{e}^{\bar m/T_\mathrm{S}} \underbrace{\Biggl( \mathrm{e}^{- \bar m/T_\mathrm{S}} \frac{1- \mathrm{e}^{-(\omega + q \Phi_\mathrm{H} + p \bar \Phi_\mathrm{H})/T_\mathrm{H}}}{ 1+ \mathrm{e}^{- \bar m/T_\mathrm{S}} \mathrm{e}^{-(\omega + q \Phi_\mathrm{H} + p \bar \Phi_\mathrm{H})/T_\mathrm{H}}} \Biggr)}_\textrm{Schwinger effect in Rindler$_2$} \Biggr\}. \label{dyon-ther-int}
\end{eqnarray}
The Schwinger temperatures, $T_\mathrm{S}$ and $\bar T_\mathrm{S}$, are given in Eq.~(\ref{eff tem}) with a generalized $\kappa$ in Eq.~(\ref{dkappa}). It should be noted that the mean number (\ref{dyon-ther-int}) for Schwinger pair production has the universal form as Eq.~(\ref{rn-ther-int}) for near-extremal RN black holes: the factorization of Schwinger effect into an AdS$_2$ and a two-dimensional Rindler space. The charge emission via Hawking radiation is given by the Hawking temperature with chemical potentials for electric and magnetic charges, which is intertangled with the Schwinger term.

\section{Conclusion}\label{con}

We have studied the Schwinger effect from (near-) extremal RN black holes with electric and/or magnetic charges in the (A)dS space. It is found that the asymptotic (A)dS boundary drastically changes the AdS$_2$ and $S^2$ radii of near-horizon geometry and the effective temperature for Schwinger effect and this boundary effect increases the Schwinger pair production for the AdS space but decreases it for the dS space. The smaller the radius, $L$, of AdS/dS space is, the larger the enhancement/suppression of the Schwinger effect is. A physical reasoning for this phenomenon is that the AdS asymptotic boundary pushes the event horizon inward and strengthens the electric field on it while the dS boundary pulls the event horizon toward the cosmological horizon and weakens the field on the event horizon. One interesting cosmological implication is that dyonic (near-) extremal RN or KN black holes have larger sizes and longer life times in the dS space than those in the asymptotically flat space and these primordial black holes may be a candidate for dark matter (for constraints for primordial black holes, see Ref.~\cite{Carr:2009jm}) and their binaries may provide a source for gravitational waves~\cite{Liu:2020cds}. This issue and other cosmological and astrophysical implications will be addressed in a separate paper.

\acknowledgments
S.P.K. would like thank the warm hospitality at National Central University.
The work of C.M.C. was supported by the Ministry of Science and Technology of the R.O.C. under the grant MOST 108-2112-M-008-007.
The work of S.P.K. was supported in part by National Research Foundation of Korea (NRF) funded by the Ministry of Education (2019R1I1A3A01063183).

\end{document}